\documentstyle[11pt,aaspp4,psfig]{article}
\doublespace

\begin{document}

\title{The yellow hypergiants HR\,8752 and $\rho$ Cassiopeiae 
near the evolutionary border of instability\footnote{
Based on observations  obtained with the William Herschel
Telescope, operated on the island  of La Palma by the Isaac Newton 
Group in the Spanish Observatorio  del Roque de los Muchachos 
of the Instituto de Astrof\'\i sica de  Canarias.}}

\author{G.~Israelian}
\affil{Instituto de Astrof\'\i sica de Canarias, E-38200
La Laguna, Tenerife, Spain} 
\author{A.~Lobel}
\affil{Harvard-Smithsonian Center for Astrophysics,
60 Garden Street, Cambridge, MA 02138, USA}
\author{M.R. Schmidt}
\affil{N. Copernicus Astronomical Center,
PL-87-100 Toru{\'n}, ul. Rabia{\'n}ska 8, Poland}

\authoraddr{Instituto de Astrof\'\i sica de Canarias, E-38200 La Laguna, 
Tenerife, SPAIN}

\authoremail{gil@ll.iac.es, alobel@cfassp21.harvard.edu 
and schmidt@ncac.torun.pl}

\begin{abstract}

High-resolution near-ultraviolet spectra of the yellow 
hypergiants HR\,8752 and $\rho$ Cassiopeiae indicate 
high effective temperatures placing both stars near the 
low-$T_{\rm eff}$ border of the ``yellow evolutionary void''. 
At present, the temperature of HR\,8752 is higher than ever. For this
star we found $T_{\rm eff}$=7900$\pm$200 K,  whereas $\rho$ Cassiopeiae
has $T_{\rm eff}$=7300$\pm$200 K. Both, HR\,8752 and $\rho$ Cassiopeiae
have developed strong stellar winds with $V_{\infty} \simeq $ 120 
${\rm km}~{\rm s}^{-1}$ and $V_{\infty} \simeq$ 100~${\rm km}~{\rm s}^{-1}$,
respectively. For HR\,8752 we estimate an upper limit for the 
spherically symmetric mass-loss of 6.7\,10$^{-6}$M$_{\odot}$ yr$^{-1}$. 
Over the past decades two yellow hypergiants appear to have approached an 
evolutionary phase, which has never been observed before.
We present the first spectroscopic evidence of the blueward motion
of a cool super/hypergiant on the HR diagram.

\end{abstract}

\section{Introduction}

Hypergiants are supergiant stars with strongly developed large-scale 
atmospheric velocity fields, excessive mass loss and extended 
circumstellar envelopes. They are rare objects, only 
12 of them being known in our Galaxy. They are very luminous but are not 
neccessarily the most luminous objects in their spectral class. The yellow
hypergiants and their characteristics have been reviewed recently 
by de Jager (1998).
There are indications (relatively small mass; overabundance of Na and N
with respect to the Sun) that yellow hypergiants are
evolved stars, evolving from the red supergiant phase to the blue phase.
Stellar evolutionary computations (e.g. Maeder \& Meynet 1988)
place cool hypergiants in a certain area on the H-R diagram 
(3.6 $< \log  T_{eff} <$3.9, 5.3$< \log L/L_{\odot} <$5.9) and 
predict that redward loops down to 4000$\pm$1000 K occur only for
stars with $M_{\rm ZAMS} \leq$ 60 $M_\odot$.  Once in the red 
supergiant phase ($T_{eff} \sim$ 3000--4000 K), stars with 
$M_{\rm ZAMS} \geq$ 10 $M_\odot$ shrink again and evolve to become
blue supergiants. However, B$\ddot{\rm o}$hm-Vitense (1958) 
has noted that stars with $T_{\rm eff}$ near 9000 K have density 
inversions, which may indicate instability. This has led to research
on the {\em yellow evolutionary void}; the area on the H-R diagram
which occupies the region 3.8 $< \log  T_{eff} <$4.0, 
5.2$< \log L/L_{\odot} <$5.9 (de Jager \& Nieuwenhuijzen 1997). 
The physics of this ``forbidden'' region for massive evolved stars 
on their blueward evolutionary loop has been studied recently by  
Nieuwenhuijzen \& de Jager (1995) and de Jager \& Nieuwenhuijzen (1997). 
Inside the void the atmospheres are moderately unstable, which is shown 
in various ways. The atmospheres have a negative density gradient at a 
certain depth level, the sonic point of the stellar wind is situated
in photospheric levels, and the sum of all accellerations is
directed outwards during part of the pulsational cycle (Nieuwenhuijzen
\& de Jager 1995). It is expected that stars, when
approaching the void during their blueward evolution, may show signs of
instability, but the very process of approaching the void has
not yet been studied. This is a field where $no$ observations 
have guided theory so far. A monitoring of stars
approaching the void will help to understand the nature of
the instabilities, the hydrodynamics of unstable atmospheres and 
finally to answer the most important question of whether or not
these stars can pass the void.

It is believed that the Galactic hypergiants HR\,8752, $\rho$ Cas 
and IRC+10420 are presently ``bouncing'' against the ``yellow
evolutionary void'' (de Jager 1998) at $\sim$ 7500$\pm$500 K, 
while there were periods when they had $T_{eff} \sim$ 4000 K. 
The brightness of IRC+10420 in $V$-band increased by 1 mag from 
1930 to 1970 (Jones et al. 1992) and its $T_{\rm eff}$ has 
increased by 1000 K over the last 20 years (Oudmaijer et al. 1996).
We do not know how rapidly they change their $T_{\rm eff}$ but 
there are some reasons to believe that these changes are 
accompanied by variations in the mass loss (de Jager 1998).
Other hypergiants that appear to have a similar position on the HR diagram
are Var A in M33 and V382 Car (Humphreys 1978). Another interesting
object; HD\,33579 appears to be located inside the void evolving to 
the red (Humphreys et al. 1991). The maximum $T_{\rm eff}$ ever observed in 
HR\,8752 is 7170 K (de Jager 1999, private communication).
Previous ground-based spectroscopic observations of HR\,8752 and $\rho$ Cas
have been carried out only in the optical and near IR region (4000--9000 \AA). 
High-resolution {\it IUE} spectra of $\rho$ Cas and HR\,8752 have been 
discussed by Lobel et al. (1998) and  Stickland \& Lambert (1981),
respectively. In this letter we report first observations of these 
hypergiants in the near ultraviolet and communicate for the first time 
the finding of spectroscopically recorded large changes of 
the effective temperature of the cool hypergiant HR\,8752 
which cannot be ascribed to the regular variability of a 
supergiant atmosphere. This finding is based on 
a unique combination of high-resolution optical spectra 
which span a period of about 30 years. Thus, HR\,8752 turned to 
be the first cool supergiant that showed the effects of stellar 
evolution from a study of its 30 years old spectrosopcic history.

\section{The Observations}

The observations were carried out in 1998 August 4
using the Utrecht Echelle Spectrograph (UES) at the Nasmyth
focus of the 4.2-m WHT at the ORM (La Palma). 
Two spectral images of $\rho$ Cas and one
image of HR\,8752 were obtained. 
A UV-sensitive CCD detector EEV 42 4200$\times$2148 
(pixel size: 13.5$\times$13.5 $\mu$m) with 60\% quantum 
efficiency at 3200 \AA\ provided superb sensitivity down to the
atmospheric cut-off at 3050 \AA. We obtained spectra which
cover the wavelength range between 3050 and 3920 \AA\
in 40 orders at a spectral resolving power of 
$R=\lambda /\Delta\lambda\sim 55,000$. For the data reduction 
we used standard {\sc iraf} \footnote{{\sc iraf} is
distributed by the National Optical Astronomical Observatories, which is
operated by the Association of Universities for Research in Astronomy, Inc.,
under contract with the National Science Foundation, USA.} procedures.
The wavelength calibration was performed with a Th--Ar lamp.
The  final signal-to-noise (S/N) ratio varies for
the different echelle orders,  being in the range 80--160 for both stars.
Additional high resolution spectra of these stars in the 
wavelength range 3500-11\,000 \AA\ were acquired 
with SOFIN echelle spectrograph at the 2.5-m NOT (La Palma) in 1998 
October 9-10. The archival spectra from 1969 Sep. 7, 
1976 July 15 and 1978 August 8 were
obtained at the Dominion Astronomical Observatory, Victoria, Canada using
the 1.2 m telescope in the coude focus (Smoli{\'n}ski et al. 1994).
The dispersion of the spectrograms was about 6 \AA/mm and signal-to-noise 
ratio at the level of 30 to 50.

\section{Analysis and Conclusions}

Simple comparison of the near-UV spectra of HR\,8752 and $\rho$ Cas
shows that these stars are no longer spectroscopic ``twins''. It is enough 
to overplot their spectra and to identify a number of lines in
order to be convinced that the atmosphere of HR\,8752 is hotter
than that of $\rho$ Cas. Most of the absorption lines in
this spectral range belong to $\alpha$- and Fe-group elements 
Ti, Si, Cr, Sc, Fe, Mn and V. In fact, the spectrum of HR\,8752
is considerably $clean$ from blends compared with $\rho$ Cas
because of a displacement of the ionization equilibrium. 
A clear illustration is presented
in Fig. 1, where we compare one of the near UV echelle orders 
and two unblended optical Fe\,{\sc i} lines 
(selected by Lobel et al. 1998) in our targets.
We have also found that many absorption lines in the near-UV spectrum 
of $\rho$ Cas are split. The first report of this phenomenon 
dates back to Bidelman \& McKellar (1957). We confirm findings by
Sargent (1961) and Lobel (1997) that
these splits in absorption appear only in lines with 
$\chi_{\rm up} \leq$ 3 eV. Various explanations for the split
absorption cores have been suggested in the literature. The
phenomenon has been explained recently by Lobel (1997), 
showing that the line splitting is caused by static emission 
emerging from detached and cool circumstellar shells,
modelled for a fast bi-polar wind.   

In order to quantify the differences in the atmospheric conditions
of our targets, we have employed a grid of LTE, plane--parallel, 
constant flux, and blanketed model atmospheres (Kurucz 1993), computed
with {\sc atlas9} without overshooting. These models are interpolated
for several values of $T_{\rm eff}$, $\log g$. For $\rho$ Cas we
used [Fe/H]=0.3 (Lobel et al. 1998) and for HR\,8752 [Fe/H]=$-$0.5
(Schmidt 1998). Synthetic spectra were computed first, 
using the LTE code {\sc wita3} (Pavlenko 1991) which takes into account 
molecular dissociation balance  (note that our targets may have 
$T_{\rm eff}$ as low as 4000 K) and all important opacity 
sources. Atomic data were obtained from the VALD-2 database 
(Kupka et al. 1999). Our spectral window contains molecular
bands of OH, CH and NH which can be used to derive CNO abundances
and constrain the range of the atmospheric parameters. Molecular data 
for the CH (3145 \AA), NH (3360 \AA) and OH (0,0) (3120-3260 \AA) 
bands were taken from Kurucz (1993), Cottrell \& Norris (1978) 
and Israelian et al. (1998), respectively. To minimize the effects 
associated with errors in the transition probabilities of molecular 
lines, the oscillator strengths ($gf$-values) have been modified from 
their original values to match the solar atlas (Kurucz et al. 1984) 
with solar abundances (Anders \& Grevesse 1989). 
Synthetic spectra of the Sun were computed using a model with
$T_{\rm eff}$=5777 K, $\log g$=4.4, [Fe/H]=0.0, microturbulence 
$\xi=1~{\rm km}~{\rm s}^{-1}$.

Our first attempts to fit the spectral
lines located in the CH and NH regions assuming solar CNO abundances 
have shown that these molecules are simply not present in the spectra. 
We have increased the abundance of nitrogen 10 times and still found
no effect on the measured equivalent widths. This can be considered as 
clear evidence that both stars had $T_{\rm eff} >$ 6200 K (given 
the values of dissociation energies of CH, NH and OH molecules) at the
time of our observation. In fact, at $T_{\rm eff}$=6200 K we
still expect 10--20 m\AA\ lines of the OH molecule located between
3100--3200 \AA\ (Israelian et al. 1998) even if oxygen is slightly 
underabundant in $\rho$ Cas with [O/H]=$-$0.3  (Takeda \& Takeda-Hidai 1998). 
Given the S/N of the data, we could easily detect a minimum of 3-4 unblended 
OH lines if they were present in the spectra. We confirm a microturbulent 
velocity $\xi=11\pm2~{\rm km}~{\rm s}^{-1}$ in both stars (de Jager 1998).
Figure 2 shows the comparison between synthetic and observed spectra 
of both stars corresponding to the regions surrounding the CH and NH lines. 
We stress that these plots should not be considered as ``best fits''.
We only want to show basic features and blends in these regions
and  demonstrate the effect of varying $T_{\rm eff}$ on the 
synthetic spectra. We did not convolve synthetic spectra 
with Gaussian macro-broadening (which is a combined effect of 
rotation and macroturbulence) because it is not affecting the EWs 
and therefore our final values of $T_{\rm eff}$/$\log g$. 
However, we convolved them 
with a Gaussian (FWHM=0.12 {\AA}) to reproduce the instrumental profile.  
The differences between the observed and calculated equivalent 
widths have been minimized for the 
best set of $T_{\rm eff}$, $\log g$ and $\xi$ (i.e. the same 
method as used by Lobel et al. 1998). We have selected 16 spectral 
lines of Sc, Cr, Ti, etc. (Fig. 2), located in the windows
3130--3170 (near the CH band) and 3340--3380 (near the NH band) and
measured their EWs (typically 300--800 m\AA) with a multi-Gaussian
function of the {\sc splot} task of {\sc iraf}. 
The final values of the atmospheric parameters are 
$T_{\rm eff}$=7900$\pm$200 K and $\log g$=1.1$\pm0.4$ for HR\,8752 and 
$T_{\rm eff}$=7300$\pm$200 K and $\log g$=0.8$\pm0.4$ for $\rho$ Cas. 
Because of the problem with UV opacities in presently available 
models of atmospheres (such as ATLAS9) with 
$T_{\rm eff} \leq 7500$ K we think that the latter 
value can be overestimated by about 250 K
(half the amplitude of $T_{\rm eff}$ variations caused by pulsation 
determined from optical spectra), since the violet wing extensions are
not as strongly developed as was observed by us in Nov.-Dec. '93 
(see Fig. 1).

The spectrum of HR\,8752 from 1969 was analyzed with a different approach. 
Due to the limited spectral region, covering wavelengths from 
4800 till 6060 \AA, severe blending and a low signal-to-noise ratio, only
a limited number of relatively unblended lines were accessible for the 
analysis - 27 Fe\,{\sc i} and 6 Fe\,{\sc ii} lines. 
Equivalent widths were typically 
in range from 200 to 600\,m\AA. The atmospheric parameters have
been found by forcing an independence of the determined single line abundance 
on the excitation potential and the equivalent width, with a unique value of
iron abundance for both neutral and ionized lines.
 
The analysis was made using atmospheric models computed with
a modified version of the TLUSTY code. The use of ATLAS9 opacity
sources and ODF functions enables us to treat these models as an extension 
of the existing grid of ATLAS9 models (Kurucz, 1993). Both plane-parrallel and 
spherically symetric models have been calculated. For spherically 
symmetric models a luminosity value of log (L/L$_{\odot}$) = 5.50 
has been utilized, as was determined by Schmidt (1998).

The resulting parameters are $T_{eff}$=5250$\pm$250\,K, 
$\log g$=$-$0.5$\pm0.5$, [Fe/H]=$-$0.55$\pm0.25$, microturbulence
$\xi_{\mu}$=10$\pm1$\,km\,s$^{-1}$ derived for plane-parallel models, 
and $T_{eff}$=5630$\pm$200 K, $\log g$=$-$0.7$\pm0.5$, 
[Fe/H]=$-$0.46$\pm0.25$ and $\xi_{\mu}$=11$\pm1$\,km\,s$^{-1}$ with
spherically symetric models. For the latter case we compute that 
the atmospherical extension was 23 percent (being measured as the ratio 
of the geometrical distance between optical depths $10^{-4}$ and $1$ 
and the stellar radius). 

It is generally accepted that H$\alpha$ is the best indicator
for global changes in the outer part of the envelope where the
wind is accelerating in a typical cool supergiant. Variations in the 
velocity and density structure of the upper layers produce changes 
in the asymmetry of the line, while an increase of the temperature
(quasi-chromosphere)  can force the wing to go into emission. 
This effect has been clearly observed in
$\rho$ Cas (de Jager et al. 1997). However, changes in H$\alpha$
may reflect those in the chromospheric structure rather than wind variations.
For this reason it  is desirable to study wind variations in other
absorption lines. In general, winds of cool stars are subtle
and difficult to detect. Far shortward extended wings due to the wind
absorption have been observed in many Fe\,{\sc i} lines of 
$\rho$ Cas in the phase when $T_{\rm eff}$=7250 K
(Lobel et al. 1998). The upper limit of the mass-loss
rate was derived as 9.2 10$^{-5}$M$_{\odot}$ yr$^{-1}$. We have also
detected these wings in many lines in the near-UV (Fig. 3). 
In addition, we have also found violet wings extending up to 
120~${\rm km}~{\rm s}^{-1}$ in the spectrum of HR\,8752. 
Assuming $\log (L/L_{\odot})$=5.6 (de Jager 1998) and 
$\rho$=7\,10$^{-15}$ \mbox{gr cm$^{-3}$} as an upper limit of the density
for the outermost layers of the atmosphere (from the model with
$T_{\rm eff}$=7900 K and $\log g$=1.1), we estimate from
$\dot{M}=4\pi R_{*} \rho V_{\infty}$ an upper limit 
$\dot{M}_{\rm max}$=6.7 10$^{-6}$ M$_{\odot}$ yr$^{-1}$ assuming 
spherically symmetric mass loss. We derived for $\rho$ Cas 
almost the same $T_{\rm eff}$ as it had in Dec. 21 1993 
(Lobel et al. 1998). This suggests that $\rho$ Cas makes small 
``oscillations'' with an amplitide $\Delta T_{eff} \sim$ 500 K 
near the void. However, the effective temperature of HR\,8752 
has risen sharply over  the last decades and places the star 
on the border of the void. When deriving the mass loss rates we
have assumed spherically symmetric outflow. However,
one should keep in mind that the real distribution of the matter
around these hypergiants is very complex and asymmtertic
(Lobel 1997, Petrov \& Herbig 1992, Humphreys et al. 1997).

It is very unlikely that the high effective temperture of HR 8752 is
due to the extra heating produced by the secondary B1,
which is located at 200 AU from the primary
and has an orbital period of 500 yr (Piters et al. 1988).
In that case, the overall spectrum
of HR 8752 would be a combination of spectral lines formed in the hot upper
layers (heated by the secondary) and cool inner layers (Piters et al. 1988). 
This is not observed and the spectrum is quite ``normal'',  
as one would expect for a hypergiant with $T_{\rm eff} \sim 8000$ K.

Understanding the final stages of stellar evolution of stars
with 10 $M_\odot \leq$ $M_{\rm ZAMS} \leq$ 60 $M_\odot$ requires
detailed knowledge of the atmospheric pulsations and 
mass-loss mechanisms of cool hypergiants.
We are inclined to think that the large variations with 
$\Delta T_{\rm eff} \sim$ 3000--4000 K are not caused by  
pulsations but reflect some complex $evolutionary$ changes due to
the active reconstruction of the stellar interior. 
Just how enhanced mass loss occurs at bouncing, is not known.
It seems significant that a number of stars moving to the blue is clustering
at the low-temperature side of the void while none of them occurs
inside the void. This leads to the hypothesis that when approaching
the border of that area, the star may show excessive mass loss and
the development of an envelope, associated with a reduction of the
effective temperature. How frequently (maybe just once?) 
this will happen before the star eventually passes through the 
void is an open question. It is quite possible that the final passage 
of the most massive stars through the void never takes place and that
these stars finally explode as Type II supernovae.

\vspace{0.1cm}

\begin{acknowledgements} 
We thank C. de Jager and H. Nieuwenhuijzen for many discussions 
and Ilya Ilyin for helping with the reduction of the SOFIN spectra. 
We also thank the anonymous referee for the useful comments.
\end{acknowledgements}

%\end{document}

\clearpage

%\begin{figure}[hb]
\begin{figure}
\psfig{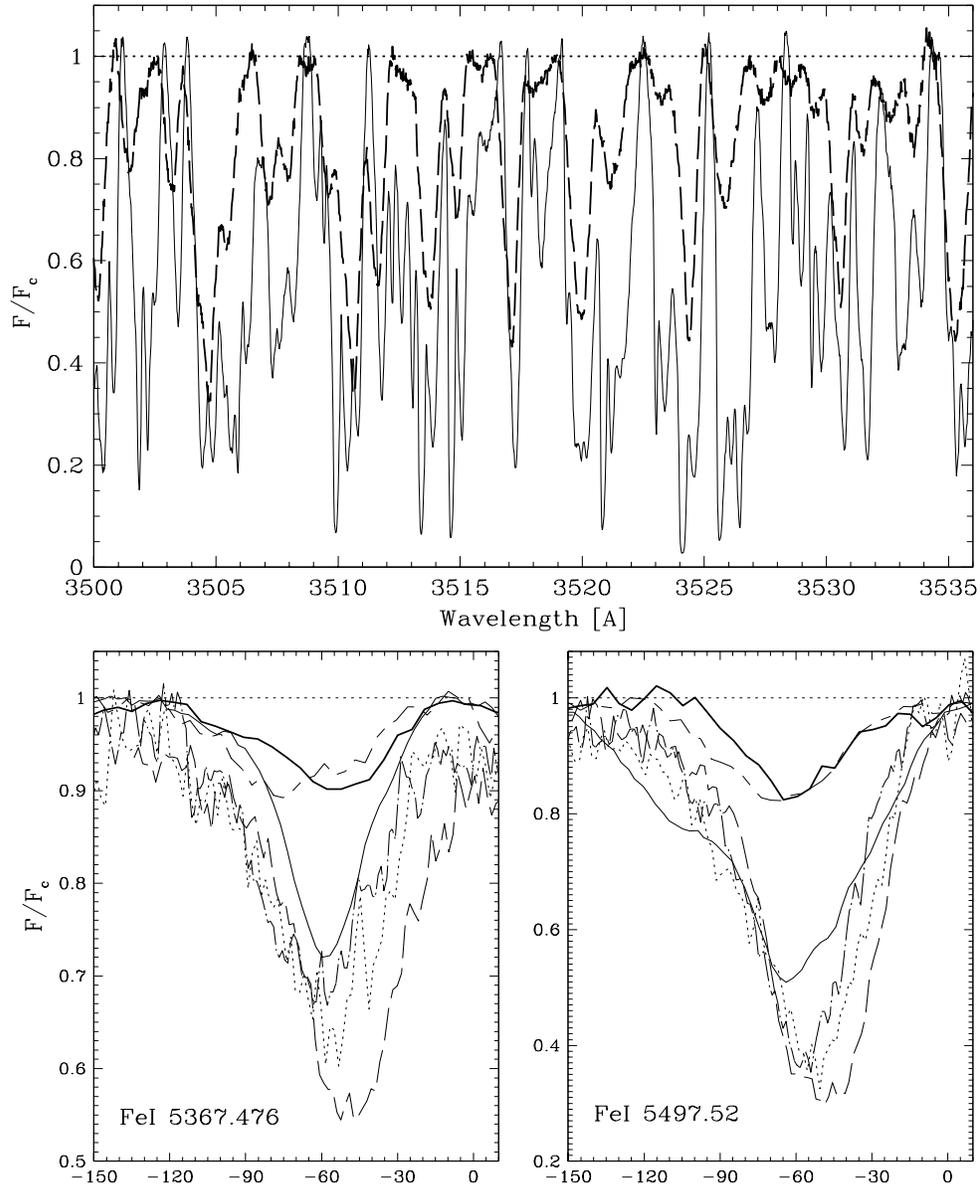}
%\figurenum{1}
%\epsscale{0.95}
%\plotone{Fig1.eps}
\caption[]{
{\bf upper panel:} 
High resolution near-UV spectra of
HR\,8752 (bold line) and of $\rho$ Cas (thin line) observed on
Aug. 4 '98 with UES. Note the single absorption cores in HR\,8752 
which appear to split in $\rho$ Cas. {\bf lower panels:} Two 
unblended Fe{\sc I} lines in both stars. The lines of HR\,8752 
developed violet wing extensions (bold line: NOT Oct. 1998, 
short dashed line: UES April 1995), 
which was also observed in $\rho$ Cas in Nov.-Dec. 1993 
when its $T_{\rm eff}$=7250 K (thin line: UES). Note the strong 
weakening of these neutral lines over the past three decades
(Dominion Obs.: long dashed line: Sept. 1969, dotted line: 
July 1975 and dash-dotted: Aug. 1978). The spectral 
resolutions are equal but note that both stars have different 
systemic velocities of $-$69 to $-$50 $\rm km s^{-1}$ and 
$-47\pm$2 $\rm km s^{-1}$ respectively.}
%\label{Fig1}}
\end{figure}

%\begin{figure}[hb]
\begin{figure}
\psfig{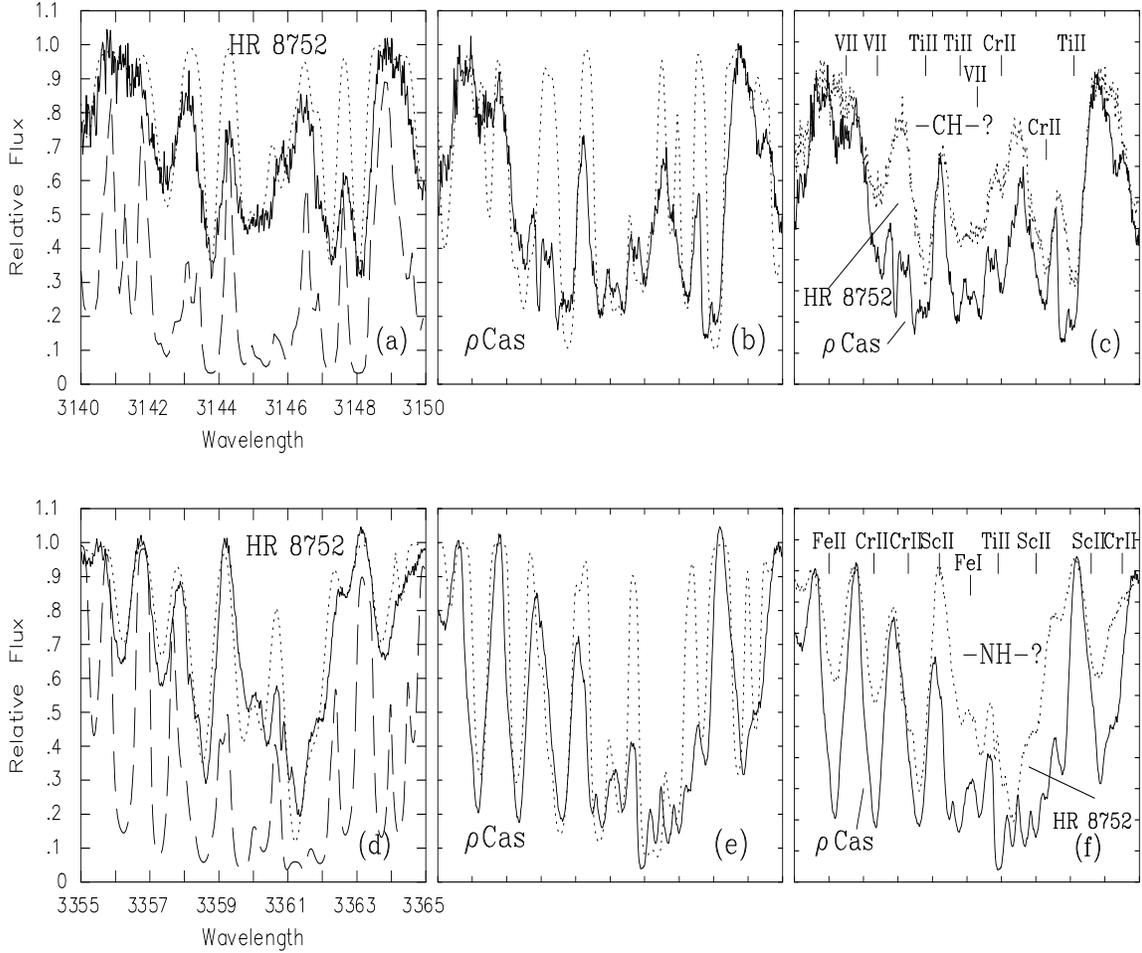}
%\figurenum{2}
%\epsscale{0.95}
%\plotone{Fig2.eps}
\caption[]{Theoretical (dotted lines) and observed (solid lines)
spectra of the regions near the CH (top row) and NH (bottom row) 
molecular bands. Theoretical spectra are computed for $T_{\rm eff}$=8100 K, 
$\log g$=1.0 for HR\,8752 (panels (a) and (d)) and for $T_{\rm eff}$=7500 K,
$\log g$=1.0 for $\rho$ Cas (panels (b) and (e)). These temperatures are
the upper limits obtained from our analysis. Dashed lines in panels
(a) and (d) correspond to the model $T_{\rm eff}$=6250 K, $\log g$=0.5.
In panels (c) and (f) we compare the observations of HR\,8752 and $\rho$ Cas.
To illustrate the effect of rotation we have convolved the synthetic spectrum
(dotted) in panel (d) with $v\sin i$=20~${\rm km}~{\rm s}^{-1}$. 
The observed spectra have been corrected for the system velocities.}
%\label{Fig2}}
\end{figure}

%\begin{figure}[hb]
\begin{figure}
\psfig{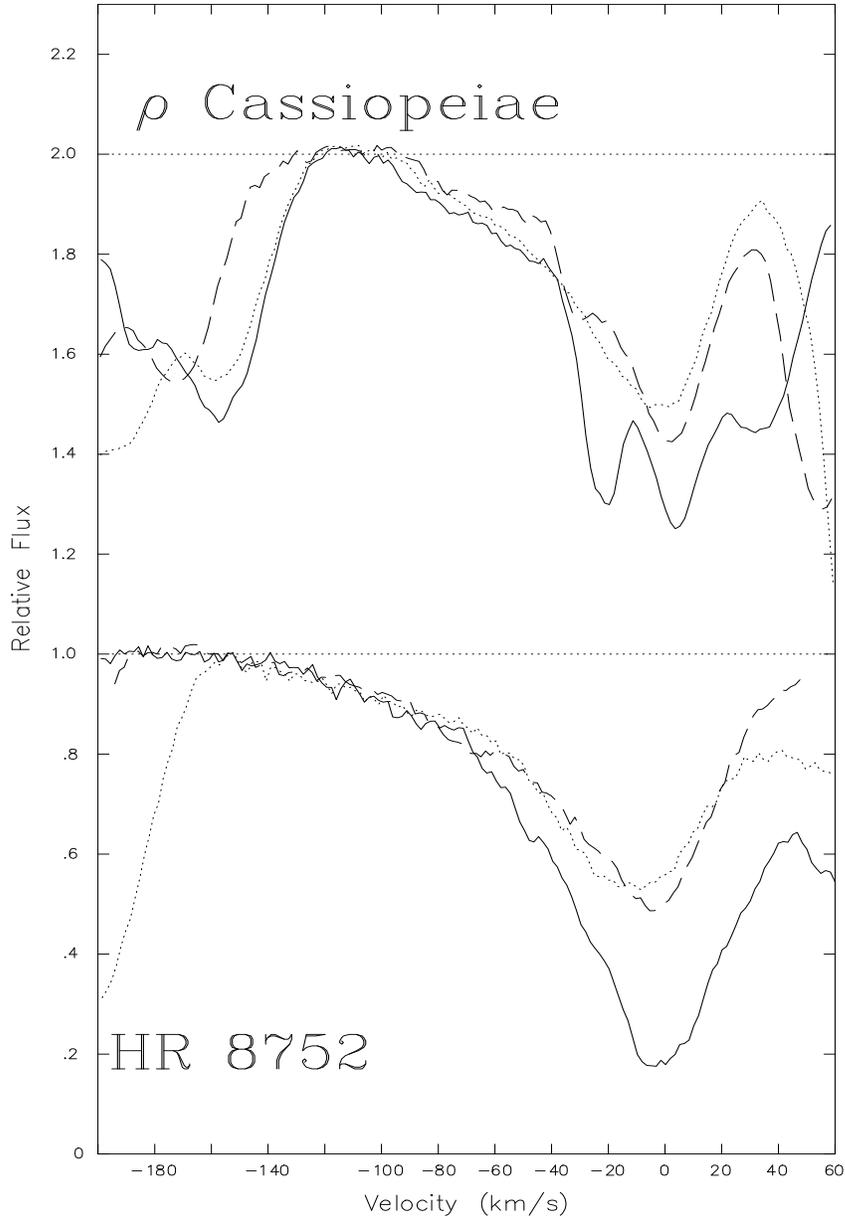}
%\figurenum{3}
%\epsscale{0.95}
%\plotone{Fig3.eps}
\caption[]{
The violet-extended wings of Fe\,{\sc ii} 3448.43
(solid line), Fe\,{\sc ii} 3436.107 (dotted line) and Fe\,{\sc i}
3640.390 \AA\ (dashed line) lines in the spectrum of $\rho$ Cas
(shifted upwards by 1.0 in upper panel) and the violet 
wings of Ti\,{\sc ii} 3335.2 (solid line),
Ti\,{\sc ii} 3500.34 (dashed line) and Si\,{\sc ii} 3862.6 \AA\ 
(dotted line) in the spectrum of HR\,8752 (lower panel). 
All lines have been shifted to their laboratory wavelengths.}  
%\label{Fig3}}
\end{figure}

\end{document}